\documentclass[letterpaper,11pt]{article}

\usepackage{epsfig}
\usepackage{graphicx}
\usepackage{color}

\usepackage{amsfonts}
\usepackage{latexsym}
\usepackage{amsmath}
\usepackage{amssymb}

\newcommand{\mikel}[1]{\marginpar{*}{\bf Mikel's remark}: {\em #1}}
\newcommand{\sidj}[1]{\marginpar{+}{\bf Sid's remark}: {\em #1}}
\newcommand{\bikd}[1]{\marginpar{=}{\bf Bikash's remark}: {\em #1}}
\newcommand{\suppress}[1]{}

\newtheorem{theorem}{Theorem}
\newtheorem{lemma}{Lemma}[section]
\newtheorem{claim}{Claim}[section]

\newtheorem{remark}{Remark}[section]

\newcommand{\bydef}{\stackrel{\triangle}{=}}

\newcommand{\poly}{\mbox{poly}}

\newcommand{\charac}{{p}}
\newcommand{\hr}{{\hat{r}}}
\newcommand{\hw}{{\hat{w}}}
\newcommand{\hW}{{\hat{W}}}
\newcommand{\hsigma}{{\hat{\sigma}}}

\newcommand{\mess}{u}

\def\QED{\mbox{\rule[0pt]{1.5ex}{1.5ex}}}

\newcommand{\F}{\mathbb{F}}
\newcommand{\Ca}{{C}}
\newcommand{\Ov}{{\tt{overwrite}}}

\newcommand{\Ad}{{\tt{additive}}}
\newcommand{\Om}{{\tt{omniscient}}}
\newcommand{\Jl}{{\tt{jam-or-listen}}}
\newcommand{\ov}{{\tt{ow}}}
\newcommand{\ad}{{\tt{add}}}
\newcommand{\om}{{\tt{omni}}}
\newcommand{\jl}{{\tt{jl}}}
\newcommand{\redun}{\bl}

\def\conf{\mbox{$\cal{S}$}}

\def\cH{\mbox{$\cal{H}$}}

\def\cK{\mbox{$\cal{K}$}}

\def\cI{{\mbox{${\cal{I}}$}}}
\def\cL{\mbox{$\cal{L}$}}

\def\cR{\mbox{$\cal{R}$}}

\def\e{\varepsilon}

\def\bx{{\bf x}}
\def\by{{\bf y}}

\def\br{{r}}
\def\Br{{\bf r}}

\def\bl{{n}}
\def\rate{{R}}
\newcommand{\Graph}{{{\cal G}}}
\newcommand{\Comp}{{{\cal C}}}

\def\01{\{0,1\}}

\newcommand{\remove}[1]{}

\begin{document}

\title{Codes against Online Adversaries}
\author{B. K. Dey\thanks{Department of Electrical Engineering,
Indian Institute of Technology Bombay, Mumbai, India, 400 076, email: {\tt bikash@ee.iitb.ac.in}}
\and S. Jaggi\thanks{Department of Information Engineering, Chinese University
of Hong Kong, Shatin, N.T., Hong Kong, email: {\tt jaggi@ie.cuhk.edu.hk}}
\and M. Langberg\thanks{Computer Science Division, Open University of
Israel, 108 Ravutski St., Raanana 43107, Israel, email: {\tt mikel@openu.ac.il}}}

\maketitle
%

\setcounter{page}{0}
\begin{abstract}
In this work we consider the communication of information in the presence of an
{\em online} adversarial jammer.
In the setting under study, a sender wishes to communicate a message to a receiver by transmitting a codeword $\bx=(x_1,\dots,x_\bl)$ symbol-by-symbol over a communication channel. The adversarial jammer can view the transmitted symbols $x_i$ one at a time, and can change up to a $p$-fraction of them.
However, the decisions of the jammer must be made in an {\em online} or {\em causal} manner.
Namely, for each symbol $x_i$ the jammer's decision on whether to corrupt it or not (and on how to change it) must depend only on $x_j$ for $j \leq  i$. This is in contrast to the ``classical'' adversarial jammer which may base its decisions on its complete knowledge of $\bx$.
More generally, for a {\em delay} parameter $d \in (0,1)$, we study the scenario in which the jammer's decision on the corruption of $x_i$ must depend solely on $x_j$ for $j \leq i - d\bl$.

In this work, we initiate the study of codes for online adversaries, and present a {\em tight} characterization of the amount of information one can transmit in both the $0$-delay and, more generally, the $d$-delay online setting.
We show that for $0$-delay adversaries, the achievable rate asymptotically equals that of the classical adversarial model. 
For positive values of $d$ we show that the achievable rate can be significantly greater than that of the classical model.

We prove tight results for both {\em additive} and {\em overwrite} jammers when the transmitted symbols are assumed to be over a sufficiently large field $\F$. In the additive case the jammer may corrupt information $x_i \in \F$ by adding onto it a corresponding error $e_i \in \F$.
In this case the receiver gets the symbol $y_i=x_i+e_i$.
In the overwrite case, the jammer may corrupt information $x_i \in \F$ by replacing it with a corresponding corrupted symbol $y_i \in \F$.
For positive delay $d$, symbol $x_i$ may not be known to the adversarial jammer at the time it is being corrupted, hence these two error models, and the corresponding achievable rates, are shown to differ substantially.

Finally, we extend our results to a {\em jam-or-listen} online model, where the online adversary can {\em either} jam a symbol {\em or} eavesdrop on it. This corresponds to several scenarios that arise in practice. We again provide a tight characterization of the achievable rate for several variants of this model.

The rate-regions we prove for each model are informational-theoretic in nature and hold for computationally unbounded adversaries. The rate regions are characterized by ``simple" piecewise linear functions of $p$ and $d$.
The codes we construct to attain the optimal rate for each scenario are computationally efficient. 
\end{abstract}
%


\newpage
\section{Introduction}

Consider the following adversarial communication scenario. A sender Alice wishes to transmit a message $\mess$ to a receiver Bob.
To do so, Alice encodes $\mess$ into a codeword $\bx$ and transmits it over a channel. In this work the codeword $\bx=x_1,\dots,x_\bl$ is considered to be a vector of length $\bl$ over an alphabet $\F$ of size $q$.
However, Calvin, a malicious adversary, can observe $\bx$ and corrupt up to a $p$-fraction of the $\bl$ transmitted symbols ({\em i.e.}, $pn$ symbols).

In the classical adversarial channel model, e.g., \cite{MS77,CT06}, it is usually assumed that Calvin has full knowledge of the entire codeword $\bx$, and based on this knowledge (together with the knowledge of the code shared by Alice and Bob) Calvin can maliciously plan what error to impose on $\bx$.
We refer to such an adversary as an {\em omniscient} adversary.
For large values of $q$ (which is the focus of this work) communication in the presence of an omniscient adversary is well-understood.
It is known that Alice can transmit no more than $(1-2p)\bl$ error-free symbols to Bob when using codewords of block length $\bl$.
Further, efficient schemes such as Reed-Solomon codes~\cite{Pet60,Ber68} are known to achieve this optimal rate.

\paragraph{Online adversaries}
In this work we initiate the analysis of coding schemes that allow communication against certain adversaries that are weaker than the omniscient adversary. We consider adversaries that behave in an {\em online} manner. Namely, for each symbol $x_i$, we assume that Calvin decides whether to change it or not (and if so, how to change it) based on the symbols $x_j$, for $j \leq i$ alone, {\em i.e.}, the symbols that he has already observed. In this case we refer to Calvin as an {\em online} adversary.

Online adversaries arise naturally in practical settings, where adversaries typically have no {\em a priori} knowledge of Alice's message $\mess$. In such cases they must simultaneously learn $\mess$ based on Alice's transmissions, and jam the corresponding codeword $\bx$ accordingly. This {\em causality} assumption is reasonable for many communication channels, both wired and wireless, where Calvin is not co-located with Alice.
For example consider the scenario in which the transmission of $\bx=x_1,\dots,x_\bl$ is done during $\bl$ channel uses over time, where at time $i$ the symbol (or packet) $x_i$ is transmitted over the channel.
Calvin can only corrupt a packet when it is transmitted (and thus its error is based on its view so far).
To decode the transmitted message, Bob waits until all the packets have arrived.
As in the omniscient model, Calvin is restricted in the number of packets $p\bl$ he can corrupt.
This might be because of limited processing power, limited transmit energy, or a need to keep his location secret.

In addition to the online adversaries described above, we also consider the more general scenario in which Calvin's jamming decisions are delayed. That is,
for a delay parameter $d \in (0,1)$, Calvin's decision on the corruption of $x_i$ must depend solely on $x_j$ for $j \leq i - d\bl$. We refer to such adversaries as {\em $d$-delay online} adversaries. Such $d$-delay online adversaries correspond, for example, to the scenario in which the error transmission of the adversary is
delayed due to certain computational tasks that the adversary needs to perform. We show that the $0$-delay model (i.e., $d=0$) and the $d$-delay model for $d>0$ display different behaviour, hence we treat them separately.

\paragraph{Error model}
We consider two types of attacks by Calvin. An {\em additive} attack is one in which Calvin can add $p\bl$ error symbols $e_i$ to Alice's transmitted symbols $x_i$. Thus $y_i$, the $i$'th symbol Bob receives, equals $x_i + e_i$. Here addition is defined over the finite field $\F_q$ with $q$ elements.
An {\em overwrite} attack is one in which Calvin overwrites $p\bl$ of Alice's transmitted symbols $x_i$ by the symbols $y_i$ received by Bob\footnote{\label{fn:0delay} Note that in the $0$-delay case these two attacks are equivalent. This is because in both cases Calvin can change an $x_i$ into an arbitrary $y_i$; an additive Calvin can choose $e_i = y_i-x_i$, whereas an overwriting Calvin directly uses $y_i$.}.
These two attacks are significantly different, if we assume that at the time Calvin is corrupting $x_i$ he has no knowledge of its value -- this is exactly the positive-delay $d$ scenario.

The two attacks we study are intended to model different physical models of Calvin's jamming. For instance, in wired packet-based channels Calvin can directly replace some transmitted packets $x_i$ with some fake packets $y_i$, and therefore behave like an overwriting adversary. On the other hand in wireless networks, Bob's received signal is usually a function of both $x_i$ and the additive error $e_i$.

Lastly we consider the {\em jam-or-listen} online adversary. In this scenario, in addition to being an online adversary, if Calvin jams a symbol $x_i$ then he has no idea what value it takes. This model is again motivated by wireless transmissions, where a node can typically either transmit or receive, but not both. For this model, we consider all four combinations of $0$-delay/$d$-delay, and additive/overwrite errors.

A rate $R$ is said to be {\em achievable} 
against an adversary Calvin
if it is possible for Alice to transmit a message $\mess$ of at least $R\bl$ symbols of $\F_q$ 
over $\bl$ channel uses to Bob (with probability of decoding error going to zero
as $n \rightarrow \infty$).
The {\em capacity}, when communicating in the presence of a certain adversarial model, is defined to be the supremum of all achievable rates.
Thus, the capacity characterizes the rate achievable in the adversarial model under study.
We denote the capacity of the classical \Om~adversarial channel which can change $p\bl$ characters by $\Ca^{\om}(p)$.
We denote the capacity of the $d$-delay online adversarial channels which can change $p\bl$ characters by $\Ca_d^{\ad}(p)$ for the \Ad~error model, and $\Ca_d^{\ov}(p)$ for the \Ov~error model. For the \Jl~adversary, we denote the corresponding capacities by $\Ca_d^{\jl,\ad}(p)$ or $\Ca_d^{\jl,\ov}(p)$, depending on whether Calvin uses additive or overwrite errors.
A more detailed discussion of our definitions and notation is given in Section~\ref{sec:defns}.



\paragraph{Our results}
In this work, we initiate the study of codes for online adversaries, and present a {\em tight} characterization of the amount of information one can transmit in both the $0$-delay and, more generally, the $d$-delay online setting. To the best of our knowledge, communication in the presence of an online adversary (with or without delay) has not been explicitly addressed in the literature. Nevertheless, we note that the model of online channels, being a natural one, has been ``on the table'' for several decades and the analysis of the online channel model appears as an open question in the book of Csisz\'{a}r and Korner \cite{CK97} (in the section addressing Arbitrary Varying Channels \cite{BBT60}). Various variants of causal adversaries have been addressed in the past, for instance \cite{BBT60,JagLHE:05,SM06,Sar08,NL08} -- however the models considered therein differ significantly from ours.

At a high level, we show that for $0$-delay adversaries the achievable rate equals that of the classical ``omniscient" adversarial model. 
This may at first come as a surprise, as the online adversary is weaker than the omniscient one, and hence one may suspect that it allows a higher rate of communication.
We then show, for positive values of the delay parameter $d$, that the achievable rate can be significantly greater than those achievable against omniscient adversaries. 

We stress that our results are information-theoretic in nature and thus hold even if the adversary is computationally unbounded. The codes we construct to achieve the optimal rates are computationally efficient to design, and for Alice and Bob to implement (i.e., efficiently encodable and decodable).
All our results assume that the field size $q$ is significantly larger than $n$.
In some cases it suffices to take $q=\poly(\bl)$, but in others we need $q = \exp(\poly(\bl))$.
Both settings lend themselves naturally to real-world scenarios, as in both cases a field element $x_i$ can be represented by a polynomial (in $\bl$) number of bits.

The exact statements of our results are in Theorems~\ref{the:nodelay},~\ref{the:add},~\ref{the:over} and~\ref{the:jl} below.
The technical parameters (including rate, field size, error probability, and time complexity) of our results are summarized in Table~\ref{table:1} of the Appendix.
We start by showing that in the $0$-delay case, the capacity of the online channel equals that of the stronger omniscient channel model.

\begin{theorem}[0-delay model]
\label{the:nodelay}
For any $p \in [0,1]$, communicating against a 0-delay online adversary channel under both the \Ov\ and \Ad\ error models equals the capacity under the \Om\ model.
In particular,
\begin{equation}
\Ca_0^{\ov}(p)=\Ca_0^{\ad}(p)=\Ca^{\om}(p) = (1-2p)^+=\left \{
\begin{array}{lc}
1-2p, \mbox{ }& p \in  [0, 0.5)\\
0,\mbox{ } & p \in [0.5,1]
\end{array}
\right .
.\label{eq:la_0_del}
\end{equation}
Moreover, the capacity can be attained by an efficient encoding and decoding scheme.
\end{theorem}

Next we characterize the capacity of the $d$-delay online channel under the additive error model.

\begin{theorem}[$d$ delay with \Ad\ error model]
\label{the:add}
For any $p \in [0,1]$ the capacity $\Ca_d^{\ad}(p)$ of the $d$-delay online channel for $d>0$ under the \Ad\ error model is $1-p$.
Moreover, the capacity can be attained by an efficient encoding and decoding scheme.
\end{theorem}

We then turn to study the $d$-delay online channel under the overwrite error model. The capacity we present is at least as large as that achievable against an additive or overwrite $0$-delay adversary who changes $p\bl$ symbols. However, it is sometimes significantly lower than that achievable against an additive  $d$-delay adversary.

\begin{theorem}[$d$ delay with \Ov\ error model]
\label{the:over}
For any $p \in [0,1]$ the capacity of the $d$-delay online channel under the \Ov\ error model is
\begin{equation}
\Ca_d^{\ov}(p)=\left \{
\begin{array}{lc}
1-p, \mbox{ }& p \in  [0, 0.5), p<d\\
1-2p+d, \mbox{ }& p \in  [0, 0.5), p>d\\
0,\mbox{ } & p \in [0.5,1]
\end{array}
\right .
.\label{eq:delay_cap}
\end{equation}
Moreover, the capacity can be attained by an efficient encoding and decoding scheme.
\end{theorem}

Lastly, we show that the optimal rates achievable against a jam-or-listen online adversary equal the corresponding optimal rates achievable against an online adversary, for each of the four combinations of
$0$- or $d$-delay, and additive or overwrite attacks.

\begin{theorem}[\Jl\ model]
\label{the:jl}
For any $p$ and $d$ in $[0,1]$ the capacity of the $d$-delay online channel under the \Jl\ error model is equal to that of the $d$-delay online channel:
\begin{equation}
\Ca_d^{\jl,\ad}(p) = \Ca_d^{\ad}(p),\mbox{ } \Ca_d^{\jl,\ov}(p) = \Ca_d^{\ov}(p)
.\label{eq:dl}
\end{equation}
Moreover, the capacity can be attained by the same efficient encoding and decoding schemes as in Theorems~~\ref{the:nodelay},~\ref{the:add} and~\ref{the:over}.
\end{theorem}


\paragraph{Outline of proof techniques}
The proofs of Theorems~\ref{the:nodelay},~\ref{the:add},~\ref{the:over} and~\ref{the:jl} require obtaining several non-trivial upper and lower bounds on the capacity of the corresponding channel models.
The lower bounds are proved constructively by presenting efficient encoding and decoding schemes operating at the optimal rates of communication. The upper bounds are typically proven by presenting strategies for Calvin that result in a probability of decoding error that is strictly bounded away from zero regardless of Alice and Bob's encoding/decoding schemes.

Theorem~\ref{the:nodelay} states that communication in the presence of a $0$-delay online adversary is no easier than communicating in the presence of (the more powerful) omniscient adversary. There already exist efficient encoding and decoding schemes that allow communication at the optimal rate of $1-2p$  in the presence of an omniscient adversary~\cite{Pet60,Ber68}. Thus our contribution in this scenario is in the design of a strategy for Calvin that does not allow communication at a higher rate.
The scheme we present is fairly straightforward, and allows Calvin to enforce a probability of error of size at least  $1/4$ whenever Alice and Bob communicate at a rate higher than $1-2p$. Roughly speaking, Calvin uses a two-phase {\em wait and attack} strategy.
In the first phase (whose length depends on $p$), Calvin does not corrupt the transmitted symbols but merely eavesdrops. He is thus able to reduce his ambiguity regarding the codeword $\bx$ that Alice transmits.
In the second phase, using the knowledge of $\bx$ he has gained so far, Calvin designs an error vector to be imposed on the remaining part of the codeword that Alice is yet to transmit.

Theorem~\ref{the:add} states that for $d>0$, the capacity of the $d$-delay online channel under the additive error model is $1-p$. Note that this expression is independent of $d$. In fact, even if Calvin's attack is delayed by just a {\em single} symbol, the rate of communication achievable between Alice and Bob is strictly greater than in the corresponding scenario in Theorem~\ref{the:nodelay}!
The upper bound follows directly from the simple observation that Calvin can always add $p\bl$ random symbols from $\F_q$ to the first $p\bl$ symbols of $\bx$, and therefore the corresponding symbols received carry no information. 
The lower bound involves a non-trivial code construction. 
In a nutshell, we show a reduction between communicating over the $d$-delay online channel under the additive error model and communicating  over an {\em erasure} channel.
In an erasure channel, the receiver Bob is assumed to know which of the $p \bl$ elements of the transmitted codeword $\bx$ were corrupted by Calvin.
As one can efficiently communicate over an erasure channel with rate $1-p$, e.g., \cite{CT06}, we obtain the same rate for our online channel.
The main question in now: ``In our model, how can Bob detect that a received symbol $y_i$ was corrupted by Calvin?''
The idea is to use authentication schemes which are information theoretically secure, and lend themselves to the adversarial setting at hand.
Namely, each transmitted symbol will include some internal redundancy, a signature, which upon decoding will be authenticated.
As Calvin is a positive delay adversary, it is assumed that he is unaware of both the symbol being transmitted and its signature.
It is enough that the signature scheme we construct be resilient against such an adversary.

In Theorem~\ref{the:over} both the lower and upper bound on the capacity require novel constructions.
For the upper bound we refine the ``wait-and attack'' strategy for Calvin outlined in the discussion above on Theorem~\ref{the:nodelay}, to fit the $d$-delay scenario.
For the lower bound, we change Alice and Bob's encoding/decoding schemes, outlined in the discussion above on Theorem~\ref{the:add}, to fit the $d$-delay \Ov\ model.
Namely, as before, Alice's encoding scheme comprises of an erasure code along with a hash function used to authenticate individual symbols.
However, in general, an \Ov\ adversary is more powerful than an \Ad\ adversary.
This is because an overwriting adversary can substitute any symbol $x_i$ by a new symbol $y_i$.
Thus Calvin can choose to replace $x_i$ with a symbol $y_i$ that is a valid output of the hash function.
Hence the design of the hash function for Theorem~\ref{the:over} is more intricate than the corresponding construction in Theorem~\ref{the:add}.

Roughly speaking, in the scheme we propose for the $d$-delay \Ov\ scenario, the redundancy added to each symbol $x_i$ contains information that allows {\em pairwise} authentication (via a pairwise independent hash function).
Namely, each symbol $x_i$ contains $\bl$ signatures $\sigma_{ij}$ (one for each symbol $x_j \in \bx$).
Using these signatures, some pairs of symbols $x_i$ and $x_j$ can be mutually authenticated to check whether exactly one of them has been corrupted. (For instance, symbols $x_i$ and $x_j$ such that $|i-j| < d\bl$ can be used for mutual authentication, since when Calvin corrupts either one of them he does not yet know the value of the other.)
This allows Bob to build a {\em consistency graph} containing a vertex corresponding to each received symbol, and an edge connecting mutually consistent symbols.
Bob then analyzes certain combinatorial properties of this consistency graph to extract a maximal set of mutually consistent symbols. He finally inverts Alice's erasure code to retrieve her message.
We view Bob's efficient decoding algorithm as the main technical contribution of this work.

Lastly, Theorem~\ref{the:jl} states that a \Jl\ adversary is still as powerful as the previously described online adversaries. This is interesting because a \Jl\ adversary is in general weaker than an online adversary, since he {\em never} finds out the values of the symbols he corrupts. This theorem is a corollary of Theorems~\ref{the:nodelay},~\ref{the:add} and~\ref{the:over} as follows.
The code constructions corresponding to the lower bounds are the same as in Theorems~\ref{the:nodelay},~\ref{the:add} and~\ref{the:over}. As for the upper bounds, we note that the attacks described for Calvin in Theorems~\ref{the:nodelay},~\ref{the:add} and~\ref{the:over} actually correspond to a \Jl\ adversary, and hence are valid attacks for this scenario as well.
\paragraph{Outline}
The rest of the paper is organized as follows.
In Section~\ref{sec:defns} we present a detailed description of our adversarial models together with some notation to be used throughout our work.
In Section~\ref{sec:t2} we present the proof of Theorem~\ref{the:add}.
In Section~\ref{sec:t3} we present the main technical contribution of this work, the proof of Theorem~\ref{the:over}.
Theorem~\ref{the:nodelay}, although stated first in the Introduction, follows rather easily from the proof of Theorem~\ref{the:over} and is thus presented in Section~\ref{sec:t1} of the Appendix.
Theorem~\ref{the:jl} follows directly from Theorems~\ref{the:nodelay}, \ref{the:add}, and \ref{the:over}, and is thus presented in Section~\ref{sec:t4} of the Appendix.
Some remarks and open problems are finally given in Section~\ref{sec:conc}.
The technical parameters of our results are summarized in Table~\ref{table:1} of the Appendix.

\section{Definitions and Notation}
\label{sec:defns}
For clarity of presentation we repeat and formalize the definitions presented earlier.
Let $q$ be a power of some prime integer, and let $\F_q$ be the field of size $q$. 
Throughout this work we assume that the field size $q$ is exponential in $\poly(\bl)$ (although some of our results will only need a polynomial in $\bl$ sized $q$) and that our parameters $p$ and $d$ are constant.
For any integer $i$ let $[i]$ denote the set $\{1,\dots,i\}$. Let $R \geq 0$ be Alice's {\em rate}.
An $[\bl,\bl\rate]_q$-{\em code} is defined by Alice's encoder and Bob's corresponding decoder, as defined below.

{\bf Alice:} Alice's message $\mess$ is assumed to be an element of $[q^{\bl R}]$. In our schemes, Alice will also hold a uniformly distributed {\em secret} $\br$ which is assumed to be a number of elements (say $\ell$) of $[q]$.
Alice's secret is assumed to be unknown to {\em both} Bob and Calvin prior to transmission.
Alice's {\em encoder} is a deterministic function mapping every $(w,\br)$ in 
$[q^{\bl \rate}] \times [q]^\ell$ to a vector $\bx = (x_1,\dots x_\bl)$ 
in $\F^\bl$.

{\bf Calvin/Channel:} 
We assume that Calvin is online, namely at the time that the character $x_i$ is transmitted Calvin has the knowledge of $\{x_i\}_{i \in K_i}$.
Here  the {\em knowledge set $K_i$} is a subset of $[i]$ that is defined below according to the different jamming models we study.
Using his {\em jamming function} Calvin either replaces Alice's transmitted symbol $x_i$ in $\F_q$ with a corresponding symbol $y_i$, or adds an error $e_i$ to $x_i$ such that Bob receives $y_i = x_i+e_i$.  

In this work, Calvin's knowledge sets must satisfy the following constraints. {\em Causality/$d$-delay:} Calvin's knowledge set $K_i$ is a subset of $[i - d\bl]$. {\em Jam-or-listen:} If Calvin is a \Jl\ adversary, $K_i$ is inductively defined so that it does not contain $j \leq i$ such that $y_j \neq x_j$. That is, Calvin has no knowledge of any $x_i$ he corrupts.

Calvin's jamming function must satisfy the following constraints.
For each $i$, Calvin's jamming function, and in particular the corresponding {\em error symbol}  $e_i \in \F_q$, depends solely on the set $\{x_i\}_{i \in K_i}$, Alice's encoding scheme, and Bob's decoding scheme.
{\em Additive/Overwrite:} If Calvin is an \Ad\ adversary, $y_i = x_i + e_i$, with addition defined over $\F_q$. If Calvin is an \Ov\ adversary, $y_i = e_i$. {\em Power:} Bob's received symbol $y_i$ differs from Alice's transmitted symbol $x_i$ for at most $p\bl$ values in $[i]$.

{\bf Bob:} Bob's {\em decoder} is a (potentially) probabilistic function solely of Alice's encoder and the received vector $\by$. It maps every vector $\by = (y_1,\dots y_\bl)$ in $\F^\bl$ to an element $u'$ 
of $[q^{\bl \rate}]$.

{\bf Code parameters:} Bob is said to make a {\em decoding error} if the message he decodes $u'$ differs from that encoded by Alice, $\mess$.
The {\em probability of error}  for a given message $\mess$ is defined as the probability, over Alice's secret $\br$, Calvin's randomness, and Bob's randomness, that Bob decodes incorrectly.
The probability of error of the coding scheme is defined as the maximum over all $\mess$ of the probability of error for message $\mess$.
Note that these definitions imply that a successful decoding scheme allows a {\em worst case} promise.
Namely, it implies high success probability no matter which message $\mess$ was chosen by Alice.

The rate $\rate$ is said to be {\it achievable} if for every $\e > 0$, $\delta>0$  and every sufficiently large $\bl$ there exists a {\em computationally efficient} $[\bl,\bl(\rate-\delta)]_q$-code that allows communication with probability of error at most $\e$.
The supremum of the achievable rates is called the {\it capacity} and is denoted by $\Ca$.
We denote the capacity of the $d$-delay online adversarial channels under the \Ad\ error model by $\Ca_d^{\ad}(p)$ and under the \Ov\ error model by $\Ca_d^{\ov}(p)$. For a \Jl\ adversary we denote the corresponding capacities by $\Ca_d^{\jl,\ad}(p)$ and $\Ca_d^{\jl,\ad}(p)$.

We put no computational restrictions on Calvin. This is because our proofs are information-theoretic in nature, and are valid even for a computationally unbounded adversary. However, our schemes provide computationally efficient schemes for Alice and Bob.

\begin{remark}
We can allow Calvin to be even stronger than outlined in the model above. In particular, Calvin's jamming function can also 
depend on Alice's message $\mess$, and our Theorems and corresponding proofs are unchanged. The crucial requirement is that each of Calvin's jamming functions be independent of Alice's secret $\br$, conditioned on the symbols in the corresponding knowledge set. That is, the only information Calvin has of Alice's secret, he gleans by observing $\bx$.
\end{remark}

{\bf Packets:} For several of our code constructions (specifically those in Theorems~\ref{the:add} and~\ref{the:over}), it is conceptually and notationally convenient to view each symbol from $\F_q$ as a ``packet" of symbols from a smaller finite field $\F_{q'}$ of size $q'$ instead. In particular, we assume
$(q')^m = q$. 
Here $m$ is an integer code-design parameter to be specified later.
For a codeword $\bx = x_1,\dots,x_n$, Alice treats each symbol (or packet) $x_i$ in $\F_q$ as $m$ sub-symbols $x_{i,1}$ through $x_{i,m}$ from $\F_{q^\prime}$. Similarly, she treats her secret $\br$ as $m$ sub-symbols $r_{1}$ through $r_{m}$ from $\F_{q^\prime}$.

\section{Proof of Theorem~\ref{the:add}}
\label{sec:t2}
\suppress{
\begin{theorem}[$d$ delay with \Ad\ error model]
For any $p \in [0,1]$ the capacity $\Ca_d^{\ad}(p)$ of the $d$-delay online channel for $d>0$ under the \Ad\ error model is $1-p$.
\end{theorem}

\mikel{I actually prefer $d$ to be an integer and not a fraction between 0 and 1, I did not change}

{\em Proof:} 
}

We consider block length $\bl$ large enough so that
$d > 1/\bl$. Throughout, to simplify our presentation, we assume that expressions such as $p\bl$ or $d\bl$ are integers.
We first prove that 
$1-p$ is an upper bound on  $\Ca_d^{\ad}(p)$
by showing a ``random-add'' strategy for Calvin.
Namely, consider an adversary who chooses elements of $\F_q$ uniformly at random and adds them to the first
$p\bl$ symbols in Alice's transmissions. Thus the first $p\bl$ symbols Bob receives are uniformly distributed random elements of $\F_q$, and carry no information at all. It is not hard to verify that such an adversarial strategy allows communication between Alice and Bob at rate at most $1-p$. This concludes our discussion for the upper bound.  

We now describe how Alice and Bob achieve a rate approaching $1-p$ with computationally tractable codes. 
Alice's encoding is in two phases. In the first phase, roughly speaking, she uses an erasure code to encode the approximately $(1-p)\bl$ symbols of her message $\mess$ into an erasure-codeword ${\bf v}$ with $\bl$ symbols. 
The erasure code allows $\mess$ to be retrieved from any subset of at least $(1-p)\bl$ symbols of the erasure-codeword ${\bf v}$. 
In the second phase, Alice uses $\bl$ ``short'' random keys and corresponding hash functions to transform each symbol $v_i$ of the erasure-codeword ${\bf v}$ into the corresponding transmitted symbol $x_i$. This hash function is carefully constructed so that if Calvin (a positive-delay additive adversary) corrupts a symbol $x_i$, with high probability Bob is able to detect this in a computationally efficient manner by examining the corresponding received $y_i$. Bob's decoding scheme is also a two-phase process. In the first phase he uses the hash scheme described above to discard the symbols he detects Calvin has corrupted -- there are at most $p\bl$ such symbols. In the second phase Bob uses the remaining $(1-p)\bl$ symbols and the decoder of Alice's erasure code to retrieve her message. We assume Alice's erasure code is efficiently encodable and decodable (for instance Reed-Solomon codes~\cite{Pet60,Ber68} can be used). In what follows we give our code construction in detail.

Let $q$ be sufficiently large (to be specified explicitly later in the proof). Let $m=\bl^2+2\bl$.
As mentioned in Section~\ref{sec:defns}, Alice treats each symbol of a codeword $\bx = x_1,\dots,x_n$ as a packet, by breaking each $x_i$ into $m$ sub-symbols $x_{i,1}$ through $x_{i,m}$ from $\F_{q^\prime}$. 
She partitions $x_{i,1}$ through $x_{i,m}$ into three consecutive sequences of sub-symbols of sizes $\bl^2$, $\bl$ and $\bl$ respectively.
The sub-symbols $x_{i,1}$ through $x_{i,\bl^2}$ are denoted by the set $w_i$, and correspond to the sub-symbols of $v_i$, the $i$th symbol of the erasure-codeword ${\bf v}$ generated by Alice.
The next $\bl$ sub-symbols are denoted by the set $r_i$, and consist of Alice's secret for packet $i$, namely, $\bl$ sub-symbols chosen independently and uniformly at random from $\F_{q'}$. For each $i$, $r_i$ is chosen independently.
The final $\bl$ sub-symbols are denoted by the set $\sigma_i$, and consist of the hash (or signature) of the information $w_i$ by the function $H_{r_i}$.
Here, $H_{r_i}$ is taken from a family $\cH$ of hash functions (known to all parties in advance) to be defined shortly. 
All in all, each transmitted symbol $x_i$ of Alice consists of the tuple $(w_i,r_i,H_{r_i}(w_i))$.

We now explicitly demonstrate the construction of each $w_i$ from Alice's message $\mess$.
Alice chooses $\rate = (1-2\bl/m)(1-p)$. Thus the message $\mess$ she wishes to transmit to Bob has $m\bl R= (m-2\bl)(1-p)\bl = (1-p)\bl^3$ sub-symbols over
$\F_{q'}$. 
Alice uses an erasure code (resilient to $p\bl^3$ erasures) to transform these sub-symbols of $\mess$ into the vector ${\bf v}$ comprising of $\bl^3$ sub-symbols over
$\F_{q'}$. She then denotes consecutive blocks of $\bl^2$ sub-symbols of ${\bf v}$ by the corresponding $w_i$'s.
More specifically, $w_i$ consists of the sub-symbols in ${\bf v}$ in locations $\bl^2(i-1)$ 
through $\bl^2i -1$.

Before completing the description of Alice's encoder by describing the hash family $\cH$, we outline Bob's decoder. Bob first authenticates each received symbol $y_i = (w'_i,r'_i,\sigma'_i)$ by checking that $H_{r'_i}(w'_i)=\sigma'_i$. He then decodes using the decoding algorithm of the erasure code on the sub-symbols on $w'_i$ of all symbols $y_i$ that pass Bob's authentication test.

We now define our hash family $\cH$ and show that with high probability any corrupted symbol $y_i \ne x_i$ will not pass Bob's authentication check.
More specifically, we study only corrupted symbols $y_i \ne x_i$ for which $w'_i \ne w_i$. (If $w'_i = w_i$, the erasure decoder described above will not make an error.)
Let $e_i$ be the error imposed by Calvin in the transmission of the $i$'th packet $x_i$.
Hence for an \Ad\ adversary Calvin, $e_i$ is defined by $y_i = x_i + e_i$.
Analogously to the corresponding sub-divisions of $x_i$ and $y_i$, we decompose $e_i$ into the tuple $(\hw_i,\hr_i,\hsigma_i)$. In particular, we
define the sets $\hw_i$, $\hr_i$ and $\hsigma_i$ so to satisfy $w'_i = w_i + \hw_i$, $r'_i = r_i + \hr_i$ and $\sigma'_i = \sigma_i + \hsigma_i$ (addition is performed by element-wise addition over $\F_{q'}$ of corresponding sub-symbols in each set).
For Bob to decode correctly, the property that $y_i$ fails Bob's authentication test if $\hw_i \neq 0$ needs to be satisfied with high probability.
More formally, noting that $r_i$ is not known to Calvin and thus independent of $\hw_i$,
we need for all $i$ and all $e_i$ such that $\hw_i \neq 0$, that $\Pr_{r_i}[H_{r'_i}(w'_i)=\sigma'_i \mid H_{r_i}(w_i)=\sigma_i]$ is sufficiently small.
Or equivalently,  
$
\Pr_{r_i}[H_{r_i+\hr_i}(w_i+\hw_i)=\sigma_i+\hsigma_i \mid H_{r_i}(w_i)=\sigma_i] = \Pr_{r_i}[H_{r_i+\hr_i}(w_i+\hw_i)-H_{r_i}(w_i) = \hsigma_i]$ is sufficiently small.

To complete our proof we present our hash family $\cH$.
Recall that $w_i$ consists of $\bl^2$ sub-symbols in $\F_{q'}$.
Let $W_i$ represent $w_i$ when arranged as a $\redun \times \redun$ matrix.
Let $\Br_i$ be a column vector of $\redun$ symbols corresponding to $r_i$.
We define the value of the hash $H_{r_i}(w_i)$ as the length-$\redun$ column vector ${\bf \sigma_i}$ defined as $W_i\Br_i$.
Thus for the corresponding errors $\hw_i \ne 0,\hr_i,\hsigma_i$ defined above,
$H_{r_i+\hr_i}(w_i+\hw_i)-H_{r_i}(w_i) = \hsigma_i$ iff $(W_i+\hW_i)(\Br_i+{\bf \hr_i})-(W_i\Br_i)={\bf \hsigma_i}$. 
Here $\hW_i$ is the matrix representation of $\hw_i$ and ${\bf \hr_i}, {\bf \hsigma_i}$ correspond to $\hr_i,\hsigma_i$. 
Namely, the corrupted symbol received by Bob is authenticated
only if 
$
\hW_i \Br_i={\bf \hsigma_i}-(W_i+\hW_i){\bf \hr_i}
$.

For Calvin to corrupt Alice's transmission, we assume that $\hw_i \ne 0$ or equivalently $\hW_i \neq 0$, therefore the rank of $\hW_i$ is at least $1$.
Now, in $\hW_i \Br_i={\bf \hsigma_i}-(W_i+\hW_i){\bf \hr_i}$, the left hand side depends on $r_i$ while the right hand side does not. Hence the equation is satisfied by at most $(q')^{\redun-1}$ values for the vector $\Br_i$. Since $\Br_i$ is uniformly distributed over $(\F_{q'})^\redun$ and unknown to Calvin, the probability of a decoding error is at most $1/q' =o(n^{-1})$ if $q'$ is chosen to be $n \cdot \omega(1)$.

All in all, our communication scheme succeeds if each corrupted symbol with $\hw_i \neq 0$ fails the authentication test.
This happens with probability at least $1-n/q' = 1-o(1)$ as desired.
Taking $m =n^2+2n$ the rate of the code is $(1-o(1))(1-p)$ and the field size needed is $(q')^m = \exp(\poly(n))$.
\QED

\section{Proof of Theorem~\ref{the:over}}
\label{sec:t3}
\suppress{
\begin{theorem}[$d$ delay with \Ov\ error model]
For any $p \in [0,1]$ the capacity of the $d$-delay online channel under the \Ov\ error model is
\begin{equation}
\Ca_d^{\ov}(p)=\left \{
\begin{array}{lc}
1-p, \mbox{ }& p \in  [0, 0.5), p<d\\ 
1-2p+d, \mbox{ }& p \in  [0, 0.5), p>d\\ 
0,\mbox{ } & p \in [0.5,1]

\end{array}
\right . 
.\label{eq:delay_cap}
\end{equation}
\end{theorem}
{\em Proof:} 
}

{\bf Proof of Upper bound:} 
We start by addressing the three cases in the upper bound on the capacity  $\Ca_d^{\ov}(p)$. First, if $p<d$, Calvin corrupts the first $p\bl$ symbols uniformly at random as in the proof of Theorem~\ref{the:add} to attain an upper bound of $1-p$ on the achievable rate.
Second, if $p \geq 1/2$ and the rate $R > 0$ is positive,
Calvin picks a codeword $\bx'$ uniformly at random from Alice's codebook. With probability at least $1-q^{-\rate\bl}$,
Alice's {\em true} codeword $\bx$ is distinct from the codeword $\bx'$. Calvin then
flips an unbiased coin, and depending on the outcome he corrupts
either the first half or the second half of $\bx$. This corruption is
done by replacing the symbols of $\bx$ by the corresponding symbols of
$\bx'$. 
If indeed $\bx \neq \bx'$, Bob has no way of determining whether Alice transmitted $\bx$ or $\bx'$.
Thus, Bob's probability of
decoding incorrectly is at least $\frac{1}{2}(1-q^{- \rate\bl}) \geq 
\frac{1}{4}$ for large enough $q$ and/or $\bl$. 

Finally, if $d< p < 1/2$, we present a ``wait-and-attack'' strategy for Calvin to prove that $1-2p+d$
is an upper bound on $\Ca_d^{\om}(p)$. 
Suppose not, and that rate $\rate = 1-2p+d+\e$ is achievable for some $\e >0$. 
Then there are $q^{\rate\bl}$ possible messages in Alice's codebook.
Calvin starts by eavesdropping on, but not corrupting, the first $(\rate - \e)\bl$ symbols Alice transmits.
He then overwrites the next $d\bl$ symbols with symbols chosen uniformly at random from $\F_q$.
These $d\bl$ locations convey no information to Bob. At this point (after Alice transmits $(\rate + d - \e)\bl$ symbols), the $d$-delay Calvin only knows the value of the first $(\rate - \e)\bl$ symbols of $\bx$.
It can be verified that with probability at least $1-q^{-\e \bl /2}$ over Alice's codebook, 
after Alice's first $(\rate + d - \e)\bl$ transmitted symbols, the set $\conf$ of codewords
consistent with what Bob and Calvin have observed thus far is of size at least $q^{\e\bl/2}$. 
Calvin then picks a random $\bx'$ from $\conf$. With probability at least $1-q^{-\e \bl /2}$, $\bx' $
is distinct from Alice's $\bx$. Calvin then
flips an unbiased coin, and depending on the outcome he corrupts
either the first half or the second half of the remaining $(1-(\rate +d -\e))\bl = 2(p-d)\bl$ symbols of $\bx$. 
This corruption is done by replacing the symbols of $\bx$ by the corresponding symbols of
$\bx'$. 
If indeed $\bx \neq \bx'$, Bob has no way of determining whether 
Alice transmitted $\bx$ or $\bx'$. Thus Bob's probability (over the message set and over the choice of Calvin) of decoding incorrectly is at least $\frac{1}{2}(1-q^{-\e \bl/2})^2 \geq \frac{1}{4}$.

{\bf Proof of Lower bound:}
We now prove that the rate $\Ca_d^\ov(p)$ specified in Theorem~\ref{the:over} is indeed achievable with a computationally tractable code. The scheme we present covers all positive rates in the rate-region specified in Theorem~\ref{the:over}, {\em i.e.}, whenever $p<1/2$. In particular the rate $\rate$ of our codes equal $1-p$ if $d>p$, and equals $1-2p+d$ if $d<p$.
Our scheme follows roughly the ideas  that appear in the scheme of Section~\ref{sec:t2}.
Namely, Alice's encoding scheme comprises of an erasure code along with a hash function used for authentication.
However, in general, an \Ov\ adversary is more powerful than an \Ad\ adversary, because it can be directly shown that an overwriting adversary can substitute any symbol $x_i$ by a new symbol $y_i$ that can pass the authentication scheme used by Bob in Section~\ref{sec:t2}.
We thus propose a more elaborate authentication scheme in which each symbol $x_i$ contains information that allows for {\em pairwise} authentication with every other symbol $x_j$.

Using notation similar to that of Section~\ref{sec:t2}, let $\mess$ be the message Alice would like to transmit to Bob, and ${\bf v}=v_i,\dots,v_\bl$ be the encoding of $\mess$ via an efficiently encodable and decodable erasure code (here we use Reed-Solomon codes).
Let $q$ be sufficiently large (to be specified explicitly later in the proof). Let $m=\bl^4+2\bl^3$ (note that this is significantly larger than in Theorem~\ref{the:add}).
As mentioned in Section~\ref{sec:defns}, Alice treats each symbol of a codeword $\bx = x_1,\dots,x_n$ as a packet, by breaking each $x_i$ into $m$ sub-symbols $x_{i,1}$ through $x_{i,m}$ from $\F_{q^\prime}$. 
She partitions $x_{i,1}$ through $x_{i,m}$ into three consecutive sequences of sub-symbols of sizes $\bl^4$, $\bl^3$ and $\bl^3$ respectively.
The sub-symbols $x_{i,1}$ through $x_{i,\bl^4}$ are denoted by the set $w_i$, and correspond to the sub-symbols of $v_i$, the $i$th symbol of the erasure-codeword ${\bf v}$ generated by Alice.
The next $\bl^3$ sub-symbols are arranged into $\bl$ sets of $\bl^2$ sub-symbols each, denoted by the sets $r_{ij}$ for each $j\in [\bl]$, and consist of Alice's secret for packet $i$. That is, each $r_{ij}$ consists of $\bl^2$ sub-symbols chosen independently and uniformly at random from $\F_{q'}$. For each $i$ and $j$, $r_{ij}$ is chosen independently.
The final $\bl^3$ sub-symbols arranged into  $\bl$ sets of $\bl^2$ sub-symbols each, denoted by the sets $\sigma_{ij}$ for each $j\in [\bl]$, and consist of the pairwise hashes of the symbols $x_i$ and $x_j$.
We define $\sigma_{ij}$ to be $H_{r_{ij}}(w_j)$, where $H_{r_{ij}}$ is taken from (a slight variation to) a {\em pairwise independent} family $\cH$ (known in advance to all parties).
Namely, $\sigma_{ij}$ is the hash of the information from $x_j$ using a key from the transmitted symbol $x_i$.
All in all, each transmitted symbol $x_i$ of Alice consists of the tuple $(w_i,\{r_{ij}\}_j,\{H_{r_{ij}}(w_j)\}_j)$.
Here $j=1,\dots,\bl$.

We now explicitly demonstrate the construction of each $w_i$ from Alice's message $\mess$.
Alice chooses $\rate = (1 - (2\bl^3)/m)\Ca$, where $\Ca$ is an abbreviation of the capacity $\Ca_d^\ov(p)$ specified in Theorem~\ref{the:over}. Note that $\rate$ equals $\Ca$ asymptotically in $\bl$ and $m$.
Thus the message $\mess$ she wishes to transmit to Bob has $m\rate\bl= (m-2\bl^3)\Ca\bl = \Ca\bl^5$ sub-symbols over
$\F_{q'}$. 
Alice uses an erasure code (resilient to $(1-\Ca)\bl^5$ erasures) to transform these sub-symbols of $\mess$ into the vector ${\bf v}$ comprising of $\bl^5$ sub-symbols over
$\F_{q'}$. She then denotes consecutive blocks of $\bl^4$ sub-symbols of ${\bf v}$ by corresponding $w_i$'s.
More specifically, $w_i$ consists of the sub-symbols in ${\bf v}$ in locations $\bl^4(i-1)+1$ 
through $\bl^4i$. Here $i=1,\dots,\bl$.

The remainder of the proof is as follows.
We first discuss the property of the family $\cH$ of hash functions in use, needed for our analysis.
We then describe and analyze Bob's decoding algorithm.

As mentioned above we use a (variation to a) pairwise independent hash family $\cH=\{H_r\}$ with the property that for all $w'_j \ne w_j$, the probability over $r_{ij}$ that 
$H_{r_{ij}}(w'_j)$ equals
$H_{r_{ij}}(w_j)$ is sufficiently small.
Such functions are common in the literature (e.g., see \cite{MR95,MU05}).
In fact, we use essentially the same hashes as in Theorem~\ref{the:add}, except with different inputs and dimension. 
Namely, let $W_i$ and $W'_i$ represent $w_i$ and $w'_i$ respectively arranged as $\redun^2 \times \redun^2$ matrices. 
Let $\Br_{ij}$ be a length-$\redun^2$ column vector of  symbols corresponding to $r_{ij}$.
We define the hash $H_{r_{ij}}(w_j)$ as the column vector ${\bf \sigma_{ij}} = W_i\Br_{ij}$. Note that
$H_{r_{ij}}(w'_j) = H_{r_{ij}}(w_j)$ means that 
$W'_j\Br_{ij} = W_j\Br_{ij}$, which implies that
$(W'_j-W_j)\Br_{ij} = {\bf 0}$. But by assumption $w'_j \neq w_j$, so $W'_j \neq W_j$, and so $W'_j-W_j$ is of rank at least $1$. Thus a random $r_{ij}$ satisfies $(W'_j-W_j)\Br_{ij} = {\bf 0}$ with probability $\leq 1/q'$. 




We now define Bob's decoder. Let $x_i$, $x_j$ be two symbols transmitted by Alice, and
$y_i$, $y_j$ be the corresponding symbols received by Bob.
Consider the information $w_i$, the secret $r_{ij}$ and the hash value $\sigma_{ij}$ in $x_i$, and let
$w'_i$, $r'_{ij}$ and $\sigma'_{ij}$ be the corresponding (potentially corrupted) values
in $y_i$.
Similarly consider the components of $x_j$ and $y_j$.
Bob checks for {\em mutual consistency} between $y_i$ and $y_j$.
Namely, the pair $y_i$ and $y_j$ are said to be mutually consistent if both
$\sigma'_{ij}=H_{r'_{ij}}(w'_j)$  and $\sigma'_{ji}=H_{r'_{ji}}(w'_i)$.
Clearly, if both $y_i$ and $y_j$ are uncorrupted versions of $x_i$ and $x_j$ respectively, they are mutually consistent.
By the analysis above of $H_{r_{ij}}$, if Calvin does not know the value of $r_{ij}$, does not corrupt $x_i$ but corrupts $w_j$, then the probability over $r_{ij}$ that $y_i$ and $y_j$ are consistent is at most $1/q'$.
This is because $\sigma'_{ij}=\sigma_{ij}=H_{r_{ij}}(w_{j})$, $r'_{ij}=r_{ij}$, and w.h.p. $H_{r_{ij}}(w_j) \ne H_{r_{ij}}(w'_j)$.
We conclude:
%


\begin{lemma}
\label{lem:ow_hash}
With probability at least $1-1/q'$, the following $y_i$ and $y_j$ are mutually inconsistent.
(i) {\em Causality:} \label{lem:ow_hash1}If $i > j$, $x_i = y_i$ and $w'_j \neq w_j$.
(ii) \label{lem:ow_hash2} {\em $d$-delay:} If $|i - j| < d\bl$, and Calvin corrupts exactly one of the symbols $x_i$ and $x_j$ so that either $w_i\neq w'_i$ or $w_j\neq w'_j$.
\end{lemma}

%

%

%
%

Bob decodes via the $d$-Delay Online Overwriting Disruptive Adversary Decoding ($d$-DOODAD) Algorithm, described in detail below.
We first give a high-level overview of the three major steps of $d$-DOODAD.
Bob's first step is to test pairs of received symbols $(y_i,y_j)$ for mutual consistency. In particular he considers only pairs of symbols separated by at most $d\bl$ locations; in this event Lemma~\ref{lem:ow_hash}(ii) implies that Bob detects the corruption of exactly one of a pair of symbols with high probability.

Based on the $O(d\bl^2)$ tests in the first step, in the second step he enumerates subsets of $\{y_1,\ldots,y_\bl\}$ of received symbols as ``candidate subsets" for decoding via Alice's erasure code. In particular, each of the candidate subsets satisfies the natural property that it contains at least $(1-p)\bl$ mutually consistent $y_i$'s. 
Na\"{i}vely, this enumeration seems computationally intractable since there may be as many as ${\bl}\choose{(1-p)\bl}$ such sets. However, there is also a more intricate combinatorial property (Step 2(c) in the $d$-DOODAD algorithm below) that candidate subsets must satisfy; we discuss this property after presenting the details of the algorithm. The effect of Step 2 below is to drastically curtail the number of candidate subsets that Bob needs to consider, to at most $\bl^{p/d}$, hence ensuring that this step is still computationally tractable. 

In the third step, for each of the candidate subsets generated in the previous step, Bob uses the decoder for Alice's erasure code to generate a set of linear equations that the sub-symbols of her message $\mess$ must satisfy. 
Then we claim that any candidate subset that has even one corrupted symbol must generate a set of inconsistent linear equations. Hence Bob decodes by using the decoder for Alice's erasure code on the unique candidate subset that generates a consistent set of linear equations.
As we will see, the error probability of our scheme will be $n^2/q'$, which is $o(1)$ if we set $q = \exp(\poly(n))$.

The details of $d$-DOODAD now follow. We define a {\em connected component} $\Graph_i$ of an undirected graph $\Graph$ as a connected subgraph of $\Graph$ such that there is no edge in $\Graph$ between any vertex in $\Graph_i$ and any vertex outside it. Also, let $\cL$ be the linear transform of the Reed-Solomon code that takes the length-$\Ca\bl^5$ column vector ${\bf u}$ of Alice's message $\mess$ to the length-$\bl^5$ column vector of the erasure codeword ${\bf v}$.  
Hence $\cL {\bf u} = {\bf v}$.
Let the column vector of sub-symbols corresponding to 
${\bf v}$ in the transmission Bob receives be denoted ${\bf w'}$. For any subset $\cI \subseteq [n^5]$ of size $\Ca\bl^5$, let $\cL_\cI$, ${\bf v}_\cI$ and ${\bf w'}_\cI$ be respectively defined as the restriction of $\cL$ to the $i$th rows/indices of $\cL$, ${\bf v}$ and ${\bf w'}$ respectively, for all $i \in \cI$. \vspace{2mm}

\noindent
{\bf $d$-Delay Online Overwriting Disruptive Adversary Decoding ($d$-DOODAD) Algorithm :}
\begin{enumerate}
\item \label{step:graph}Bob constructs a {\em $d$-distance mutual consistency graph} $\Graph$ with vertex set $\{y_1,\ldots,y_\bl\}$ and edge-set comprising of all mutually consistent pairs $(y_i,y_j)$ such that $|i - j| < \bl d$ (but no other edges).
Thus $\Graph$ comprises of $\ell \leq \bl$ connected components $\{\Graph_1,\dots\Graph_\ell\}$. \vspace{-1mm}

\item \label{step:cset} Let $\cK$ be a subset of $[\ell]$.
We define the {\em candidate subset $\Comp(\cK)$ of $\Graph$} as the set $\{\Graph_k|k \in \cK\}$ of connected components in $\Graph$. If the size of $\cK$ is $j$, we say $\Comp(\cK)$ {\em has size $j$}.
Bob enumerates all possible candidate subsets $\Comp(\cK)$ of $\Graph$ such that
(a) The candidate subset $\Comp(\cK)$ has size at most $c = p/d $.
(b) The number of vertices in the subgraphs in $\Comp(\cK)$ is at least $(1-p)\bl$.
(c) Each pair of vertices $y_i$ and $y_j$ in the union of the subgraphs in $\Comp(\cK)$ are mutually consistent.\vspace{-1mm}

\item \label{step:decod}
Let $\bar{\cK} \subseteq [n^5]$ be the set comprising of indices in ${\bf w'}$ corresponding to all symbols $y_i$ in the components $\Comp(\cK)$.
Bob picks an arbitrary subset $\cI \subset \bar{\cK}$ of size $\Ca\bl^5$. 
If 
$\cL_{\bar{\cK}} \left( \left( \cL_\cI \right)^{-1}{\bf w'}_\cI\right) = {\bf w'}_{\bar{\cK}}$,
he decodes $\mess$ as the sub-symbols in the vector $\cL_\cI ^{-1}{\bf w'}_\cI$. Otherwise he discards $\cK$ and returns to the beginning of Step~\ref{step:decod}.\vspace{-1mm}

   
  
\end{enumerate}

\begin{claim} The $d$-DOODAD algorithm decodes Alice's message correctly with probability at least $1-\bl^2/q'$.
\end{claim}
{\em Proof:} Throughout we assume that Lemma~\ref{lem:ow_hash} holds for all corresponding $y_i$ and $y_j$ (by the union bound this happens with probability at least $1-\bl^2/q'$).
Thus corrupted $y_i$ and uncorrupted $y_j$ are non-adjacent  in $\Graph$. 
We first prove that at least one $\Comp(\cK)$ with only uncorrupted symbols satisfies Steps~\ref{step:cset} and~\ref{step:decod}. We examine the three conditions of Step~\ref{step:cset}. By the definition of mutual consistency any set with only uncorrupted symbols satisfies Step 2(c). 
Since Calvin can corrupt at most $p\bl$ symbols, there must be some $\Comp(\cK)$ satisfying Step 2(b). To prove that $\Comp(\cK)$ also satisfies Step 2(a), we observe the following.
If Calvin does not corrupt at least $d\bl$ consecutive symbols between two uncorrupted symbols $y_i$ and $y_j$ (say I<j),  there must be a sequence of at most $j-i+1$ uncorrupted symbols with indices $i = k_0 \leq k_1 \leq  k_2 \leq \ldots \leq k_{j-i} = j$ such that any two consecutive symbols in the sequence have indices that differ by less than $d\bl$. Then by the definition of 
$\Graph$, both $y_i$ and $y_j$ must be in the same connected component of $\Graph$. But there are at most $p \bl$ corrupted symbols, 
hence there are at most $c = p/d $ disjoint sequences of $\bl d$ consecutive corrupted symbols (and thus at most $c$ components in $\Comp(\cK)$).

Lastly, we show that any $\Comp(\cK)$ with only uncorrupted symbols and satisfying Step 2 must also satisfy Step 3. To see this, note that any such $\Comp(\cK)$ has at least $(1-p)\bl$ symbols from $\F_q$. Thus, by the definitions of $m$ and $\Ca$ for Theorem~\ref{the:over}, $\Comp(\cK)$ has at least $(1-p)\bl^5 \geq \Ca\bl^5$ uncorrupted sub-symbols over $\F_{q'}$. Also, since $\Comp(\cK)$ comprises solely of uncorrupted symbols, ${\bf w'}_{\bar{\cK}} = {\bf v}_{\bar{\cK}}$, hence for any $\cI$, ${\bf w'}_\cI = {\bf v}_\cI$. But by the properties of erasure codes, 
$\cL_\cI ^{-1}{\bf v}_\cI = {\bf u}$, Alice's message vector.
Thus $\cL_{\bar{\cK}} \left(  \cL_\cI ^{-1}{\bf w'}_\cI\right) = \cL_{\bar{\cK}}{\bf u} = {\bf v}_{\bar{\cK}} = {\bf w'}_{\bar{\cK}}$.

We now show that there does not exist any $\Comp(\cK')$ such that the corresponding output of the $d$-DOODAD algorithm $\mess(\Comp(\cK'))$ differs from Alice's real message $\mess$.
We prove this by contradiction. Suppose a $\Comp(\cK')$ passes all the decoding steps of the $d$-DOODAD algorithm and results in a $\mess(\Comp(\cK'))$ distinct from Alice's message $\mess$. We now make a series of observations that successively refine the structure of such a $\Comp(\cK')$, resulting in the conclusion that, w.h.p., $\Comp(\cK')$ contains no uncorrupted symbols, and therefore $\mess(\Comp(\cK')) = \mess$.

First, note that $\Comp(\cK')$ must contain uncorrupted symbols to
pass Step 2(b), since  $p<1/2$.
In addition, to pass Step 2(c), by Lemma~\ref{lem:ow_hash}(i), all the uncorrupted symbols of $\Comp(\cK')$ must come before all the symbols corrupted by Calvin. Now notice that the uncorrupted and the corrupted symbols in $\Comp(\cK')$ must be separated by a {\em separating set} $\cR$ of at least $\bl d$ consecutive symbols not in $\Comp(\cK')$. If not, Lemma~\ref{lem:ow_hash}(ii) would imply that w.h.p. $\Comp(\cK')$ does not satisfy Step 2(c) of $d$-DOODAD.
Now note that the separating set $\cR$ must contain at least $d\bl$ consecutive symbols corrupted by Calvin.
This follows from the fact that $\Comp(\cK')$ consists of connected components.
Namely, if $\cR$ contains less than $d\bl$ corrupted symbols, there must exist an uncorrupted symbol $y_i$ and a corrupted symbol $y_j$, both in $\Comp(\cK')$, satisfying $|j - i| < d\bl$. But this by Lemma~\ref{lem:ow_hash}(ii) would contradict Step 2(c). Notice that if $d>p$ we may conclude our proof at this point.

We now observe that there are at most $(p-d)\bl$ corrupted symbols in $\Comp(\cK')$. This follows from the fact that $\cR$ contains $d\bl$ consecutive symbols corrupted by Calvin (not in $\Comp(\cK')$), and the fact that Calvin can corrupt at most $p\bl$ symbols.
This, together with Step 2(b) of $d$-DOODAD, implies that the component set $\Comp(\cK')$ contains a proper subset $\Comp(\cK'')$  with at least $\Ca\bl$ uncorrupted symbols.
Finally, let $\cI$ be any subset of $\Ca\bl^5$ uncorrupted sub-symbols in
$\Comp(\cK'')$. Let $\cI'$ be any other subset of $\Ca\bl^5$ symbols in 
$\Comp(\cK'')$.
Consider the corresponding message vectors 
${\bf u} = \cL_\cI ^{-1}{\bf w'}_\cI$ and ${\bf u'} =\cL_{\cI'} ^{-1}{\bf w'}_{\cI'}$
that Step~\ref{step:decod} of $d$-DOODAD may decode to.
Since ${\bar{\cK'}}$ is of size at least $(1-p)\bl^5$, by the property of erasure codes~\cite{MS77}, if ${\bf u'} \neq {\bf u}$, then $\cL_{\bar{\cK'}}{\bf u'}\neq \cL_{\bar{\cK'}}{\bf u}$. Thus $
\cL_{\bar{\cK'}} \left( \cL_{\cI'} ^{-1}{\bf w'}_{\cI'}\right)
\neq 
\cL_{\bar{\cK'}} \left( \cL_\cI ^{-1}{\bf w'}_\cI\right) 
= \cL_{\bar{\cK'}}{\bf u}
={\bf w'}_{\bar{\cK'}}$, contradicting Step 3. \QED


\section{Conclusion}
\label{sec:conc}

In this work we characterize the capacity of online adversarial channels and their variants under the \Ad\ and \Ov\ error models.
Our results are tight and coding schemes efficient.
Throughout, we assume that the communication is over a size $q$ alphabet, assumed to be large compared to the block-length $\bl$.
An intriguing problem left untouched in this work concerns communication in the online adversarial setting over ``small", {\em e.g.} binary, alphabets.
The authentication schemes used extensively in this work depend integrally on the the alphabet size being large. They do not extend na\"{i}vely to the binary alphabet case, where new techniques seem to be needed.

\newpage
\bibliographystyle{plain}
\bibliography{online}

\appendix

\section{List of parameters of our codes}

\begin{table}[h]
\begin{tabular}{|c|c|c|c|c|c|}
\hline
 \multicolumn{2}{|c|}{} & Capacity & Minimum $q$ & Complexity & Probability of Error \\ \hline
\multicolumn{2}{|l|}{Theorem 1} & $1-2p$ & $q>\bl$ & ${\cal O}\left ( \bl^2\log \bl \log^3 q\right )$ & $0$ \\ \hline
\multicolumn{2}{|l|}{Theorem 2}  & $1-p$ &   $\bl^{\Omega(1/\delta^2)}$ & ${\cal O}\left ( \bl^2\log \bl \log ^3q\right )$ & ${\cal O}\left (  \bl q^{-\delta^2}\right )$ \\ \hline
Theorem 3 & $d<p<0.5$ & $1-2p+d$ &   $\bl^{\Omega(\bl^2/\delta^2)}$
& ${\cal O}\left ( \bl^{p/d+2}\log \bl\log ^3q \right )$ & ${\cal O}\left (  n^2q^{-\delta^2/\bl^2}\right )$ \\ \cline{2-6}
& $p<d, p<0.5$ & $1-p$ &$\bl^{\Omega(\bl^2/\delta^2)}$ & ${\cal O}\left ( \bl^2\log\bl \log^3 q  \right )$ & 
${\cal O}\left (  \bl^2q^{-\delta^2/\bl^2}\right )$ \\\hline
\end{tabular}
\caption{Bounds on the capacity $\Ca$, alphabet size $q$ required to achieve capacity, computational complexity, and probability of error, of our main results. The bounds are in terms of the parameters $p$ (adversary's power), $d$ (adversary's delay), $\bl$ (block-length), $q$ (field-size), and $\delta$ (difference between the $\Ca$ and rate $\rate$).}
\label{table:1}
\end{table}

Table~\ref{table:1} is obtained by careful analysis of the parameters of the algorithms corresponding to Theorems~\ref{the:nodelay}, \ref{the:add} and \ref{the:over}. The corresponding values for the scenarios in Theorem~\ref{the:jl} are omitted since they are element-wise identical to those in the table. The values in Table~\ref{table:1} substitute the rate-overhead parameter $\delta$ for the packet-size parameter $m$ used in the proofs of Theorems~\ref{the:add} and \ref{the:over} since we feel this choice of variables  is more ``natural" when examining the tradeoffs between code parameters. Also, the algorithms presented in the proofs of Theorems~\ref{the:add} and \ref{the:over} correspond to a particular setting of the $\delta$ parameter; we omitted this degree of freedom in the presentation of the proofs, for ease of exposition. Lastly, no effort has been made to optimize the tradeoffs between the parameters in Table~\ref{table:1}; in fact, we have preliminary results on schemes that improve on some of these parameters (work in progress).

\section{Proof of Theorem~\ref{the:nodelay}}
\label{sec:t1}


As discussed in the Introduction, the lower bound of Theorem~\ref{the:nodelay} 
follows from known constructions \cite{Pet60,Ber68}.
To complete the proof, then, all that is needed is a corresponding upper bound on the capacity.
The required upper bound is novel. However, it is a special case of upper bound of Theorem~\ref{the:over}, and follows directly if the parameter $d$ in the corresponding proof is set to zero. 

\section{Proof of Theorem~\ref{the:jl}}
\label{sec:t4}

In the \Jl\ online model, Calvin is assumed to be unaware of the value of the symbols $x_i$ that he corrupts. 
Theorem~\ref{the:jl} states that a \Jl\ adversary is still as powerful as the previously described online adversaries, and is actually a corollary of Theorems~\ref{the:nodelay},~\ref{the:add} and~\ref{the:over}.
First of all, the code constructions corresponding to the lower bounds are the same as in Theorems~\ref{the:nodelay},~\ref{the:add} and~\ref{the:over}. As for the upper bounds, it is not hard to verify that the attacks for Calvin outlined in each of the settings addressed in the paper correspond to a \Jl\ adversary, and hence are valid attacks for this scenario as well.

\end{document}